\documentclass[twocolumn, prl, aps, superscriptaddress, showpacs]{revtex4}        
\usepackage{graphicx}
\usepackage{wasysym}
\begin{document}
\title{Strain dependence of the acoustic properties of amorphous metals below 1K:\\
Evidence for the interaction between tunneling states}
\author{R. K\"onig}
\affiliation{Physikalisches Institut, Universit\"at Bayreuth, D-95440 Bayreuth, Germany}
\affiliation{Abteilung Supraleitung und Magnetismus, Universit\"at Leipzig, Linn\'estr. 5, D-04103 Leipzig, Germany}
\author{M. A. Ramos}
\affiliation{Departamento de F\'{\i}sica de la Materia Condensada, C-III, Instituto Nicol\'as Cabrera, \\
Universidad Aut\'onoma de Madrid, E-28049 Madrid, Spain}
\author{I. Usherov-Marshak}
\affiliation{Physikalisches Institut, Universit\"at Bayreuth, D-95440 Bayreuth, Germany}
\author{J. Arcas-Guijarro}
\author{A. Hernando-Ma\~neru}
\affiliation{Instituto de Magnetismo Aplicado, Las Rozas, E-28230 Madrid, Spain}
\author{P. Esquinazi}
\affiliation{Abteilung Supraleitung und Magnetismus, Universit\"at Leipzig, Linn\'estr. 5, D-04103 Leipzig, Germany}
\date{\today}

\begin{abstract}
We have conducted a thorough study of the acoustic properties between $10^{-4}$ and 1 Kelvin for the amorphous metal Zr$_{\rm x}$Cu$_{\rm 1-x}$ (x=0.3 and x=0.4), by measuring the relative change of sound velocity $\Delta v/v$ and internal friction $Q^{-1}$ as a function of temperature and also of the applied strain, in both superconducting and normal state. We have found that when plotted versus the ratio of strain energy to thermal energy, all measurements display the same behavior: a crossover from a linear regime of ``independent'' tunneling systems at very low strains and/or high enough temperatures to a nonlinear regime where $\Delta v/v$ and $Q^{-1}$ depend on applied strain and the tunneling systems cannot be considered as independent. 
\end{abstract}


\maketitle

The standard tunneling model (STM), independently introduced by Anderson, Halperin and Varma \cite{ande72}, and Phillips \cite{phil72} in 1972, provides a widely accepted basis for the phenomenological description of thermal, dielectric and acoustic properties of amorphous solids below~1 K \cite{phil_rev,hunk86,esqu98}.
The central idea of the model is the universal existence in glasses of a random distribution $P (\Delta_0 , \Delta) = P_0 / \Delta_0 $ of {\em independent} two-level states or tunneling systems (TS) with quantum energy splitting $\Delta_0$ and asymmetry $\Delta$. 
Although low-temperature thermal properties have often been used to confirm the STM predictions (especially in non-metallic glasses devoid of the free-electron contributions which can overwhelm those of TS), indeed acoustic properties in the kHz range are much more sensitive to details of the TS and their interactions with phonons and/or electrons.
At very-low temperatures and not very-high frequencies ($T \ll T_{\rm co}$), the STM predicts \cite{phil_rev,hunk86,esqu98} for {\em dielectric} glasses a relative change of sound velocity $\Delta v/v = C \ln(T/T_0)$ and an internal friction $Q^{-1} \propto C T^3 / \omega$. Above the crossover temperature $T_{\rm co}$ ($T \gg T_{\rm co}$), (phonon) relaxational processes add to resonant contributions to the sound velocity and $\Delta v/v = - C/2 \ln(T/T_0)$, whereas the relaxation-dominated internal friction reaches a constant ``plateau'' value $Q^{-1} = (\pi/2) C$. 
$C$ is the fundamental, dimensionless parameter of the STM ($C = P_0 \gamma^2 / \rho v^2$, where $\gamma$ denotes a TS-phonon coupling constant, $v$ and $\rho$ are sound velocity and density of the material) and $T_0$ is an arbitrary reference temperature. 
In amorphous {\em metals} \cite{blac81}, conduction electrons provide in addition to phonons an alternative channel for relaxation of the TS, typically dominant below 1 K. In this case, $T_{\rm co} \sim$ 1--10 $\mu $K is estimated instead of $T_{\rm co} \sim $ 100 mK as for dielectric glasses, and experiments should therefore show a sound-velocity variation with both resonant and electron relaxational contributions $\Delta v/v = C/2 \ln(T/T_0)$, and an internal friction with a wide ``plateau'' $Q^{-1} = (\pi/2) C$. The good overall agreement between these predictions and many experiments \cite{phil_rev,hunk86,esqu98,rayc84} gave the definitive support to the STM. Nevertheless, significant discrepancies below $\sim 100 \,$mK have also been reported: internal friction in dielectric glasses decreases for $T < T_{\rm co}$ much more slowly than $Q^{-1} \propto T^3$ \cite{esqu98,esqu92,hunk00}, the slopes in $\Delta v/v$ are not always as predicted, and the acoustic properties of both dielectric and metallic glasses exhibit a strong, unexpected strain dependence \cite{esqu92,esqu98,ramos00}.

On the other hand, Yu and Leggett \cite{yu} have questioned the general idea of independent, noninteracting TS in glasses and the very STM, arguing that the universality of low-energy excitations may be the result of interactions between some kind of defects. They emphasized that the most important universality of glasses (compatible with, but not explained by the STM) is that at low frequencies $ l \sim 150 \lambda $ as found by Freeman and Anderson \cite{freeman} ($l$ and $\lambda$ being phonon mean free path and wavelength, respectively), and hence $Q^{-1}_{\rm plateau} \approx P_0 \gamma^2 / \rho v^2 \sim 10^{-3}$, despite a great variation in $P_0 , \, \gamma , \, \rho$ and $v$. On general grounds \cite{yu},  the effective interaction between two TS separated by a distance $r$ is dipolar elastic and the interaction energy reads $\sim U_0 / r^3$, with $U_0 \approx \gamma^2 / \rho v^2$. The observed universality of low-temperature glassy properties can therefore be written in terms of the dimensionless interaction strength as $ P_0  U_0 \approx {\rm const} \sim 10^{-3}$ (for a review, see Burin {\it et al.} \cite{burin}). In practice, $ P_0  U_0 $ could be a few times larger than experimental data of $Q^{-1}_{\rm plateau}$, since the former is mainly determined by transverse phonons whereas the latter usually measures the absorption of longitudinal sound waves $ P_0 \gamma_{l}^2 / \rho v_{l}^2$ \cite{justify}.

Although several attempts have been made to understand aforementioned deviations from the STM of internal friction and dielectric absorption by interacting-TS induced relaxations \cite{burin}, no consensus in this matter has been achieved. Our aim here has been to address these open questions, by studying the acoustic properties of amorphous Zr$_{30}$Cu$_{70}$ as a function of temperature {\em and} of applied strain in the range 0.1~mK $ < T < $ 1~K, both in the superconducting ($T_{\rm c} = 95$~mK) and in the normal state, therefore being able to tune phonons or electrons as TS relaxational centers. For comparison, we have also measured Zr$_{40}$Cu$_{60}$ ($T_{\rm c} = 280$~mK), for which there exist published data above 10~mK \cite{rayc84}.

Ribbons of Zr$_{\rm x}$Cu$_{\rm 1-x}$ (30 $\mu $m thick) were prepared with the melt-spinning technique in an argon-controlled environment. X-ray diffraction showed the samples were fully amorphous. Critical superconducting temperatures and magnetic fields were determined from electric resistance measurements. Young-modulus sound velocity dispersion and absorption were measured at around 1 kHz using the vibrating-reed technique in a nuclear adiabatic demagnetization cryostat.  

\begin{figure}[ht]
\includegraphics[width=7.9cm,clip]{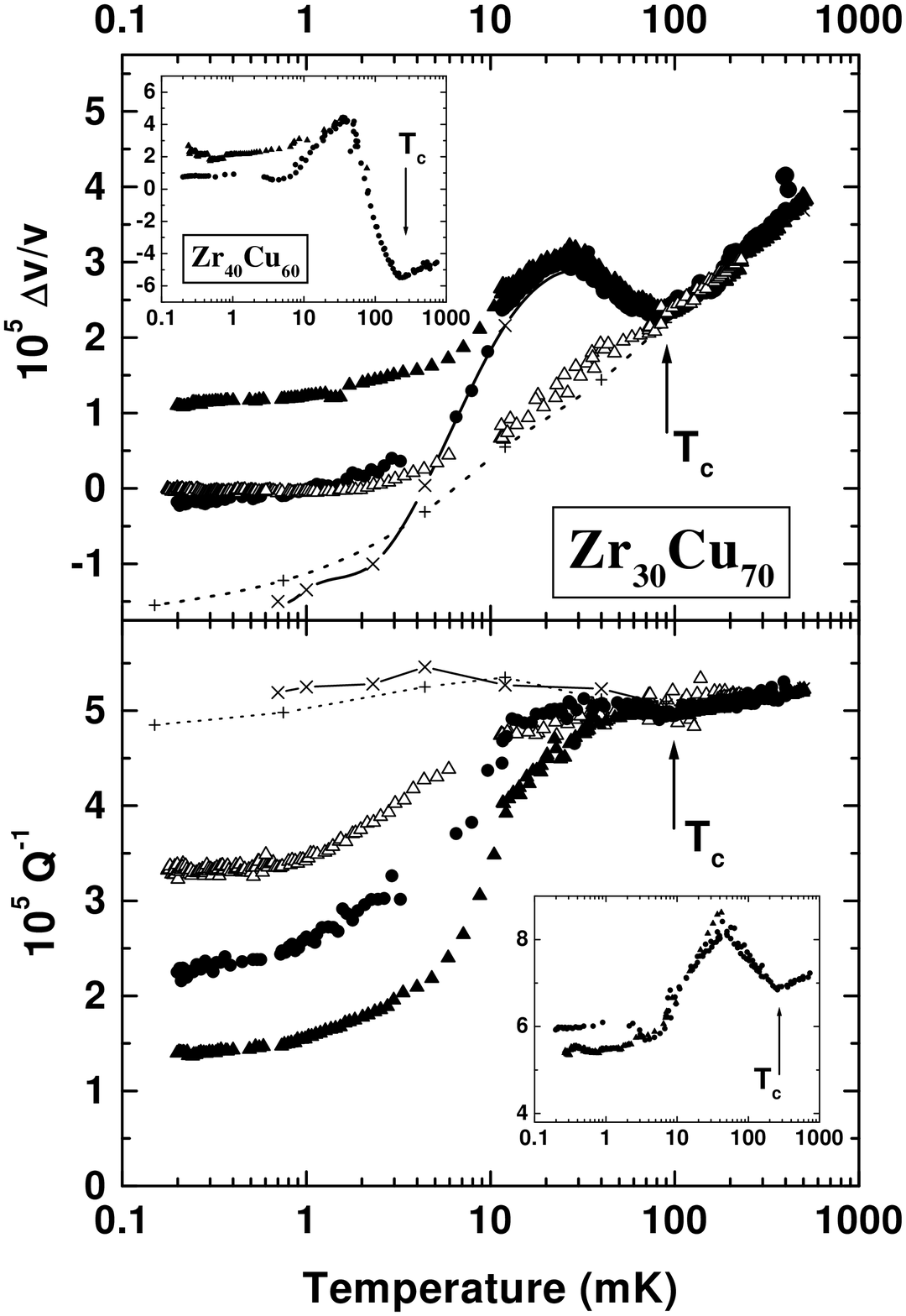}
\caption{Sound-velocity variation (top) and internal friction (bottom) for Zr$_{30}$Cu$_{70}$, measured at 1 kHz in superconducting (SC) state below 95 mK (solid symbols, $B = 0$) and in normal (N) state (open symbols, $B = 250$ mT) at different driving voltages: 2 V (triangles), 1 V (circles). Extrapolated data to zero strain (see text) are indicated by a solid line (SC) and a dashed line (N). The inset shows data for Zr$_{40}$Cu$_{60}$. The superconducting transitions are indicated by arrows.}
\label{temperature}
\end{figure}

In Fig.~\ref{temperature}, we show $\Delta v/v$ and $Q^{-1}$ data for Zr$_{30}$Cu$_{70}$ in both superconducting and normal (with a magnetic field $B = 250$~mT) states, obtained at different excitation voltages of the reed during warm-up periods of the nuclear stage \cite{ramos00}. An eddy-current contribution to the internal friction, depending quadratically on magnetic field, was measured and subtracted. Similar data for Zr$_{40}$Cu$_{60}$ (with  $B = 0$), shown in the insets of Fig.~\ref{temperature}, agree well with published data above 10~mK \cite{rayc84}, apart from an almost factor of 2 difference in the absolute value of internal friction, which may be partially due to different contributions from the clamping of the samples. The observed behavior just below $T_{\rm c} $ is thought to arise from the strong variation in the number of quasiparticles thermally activated in that temperature region \cite{esqu98,rayc84}. The maximum in $\Delta v/v$ at $T_{\rm co} \approx 30 $~mK (40~mK) for x=0.3 (x=0.4) should correspond to the abovementioned crossover, occurring when $\omega \tau \sim 1$ ($\omega$ being the angular frequency of the acoustic wave and $\tau$ the relaxation time of the TS). As expected, this peak disappears in the normal metallic state, since this crossover shifts to orders of magnitude lower temperatures due to the much faster relaxation rate of TS by electrons than by phonons.

\begin{figure}[ht]
\includegraphics[width=5.4cm,angle=270,clip]{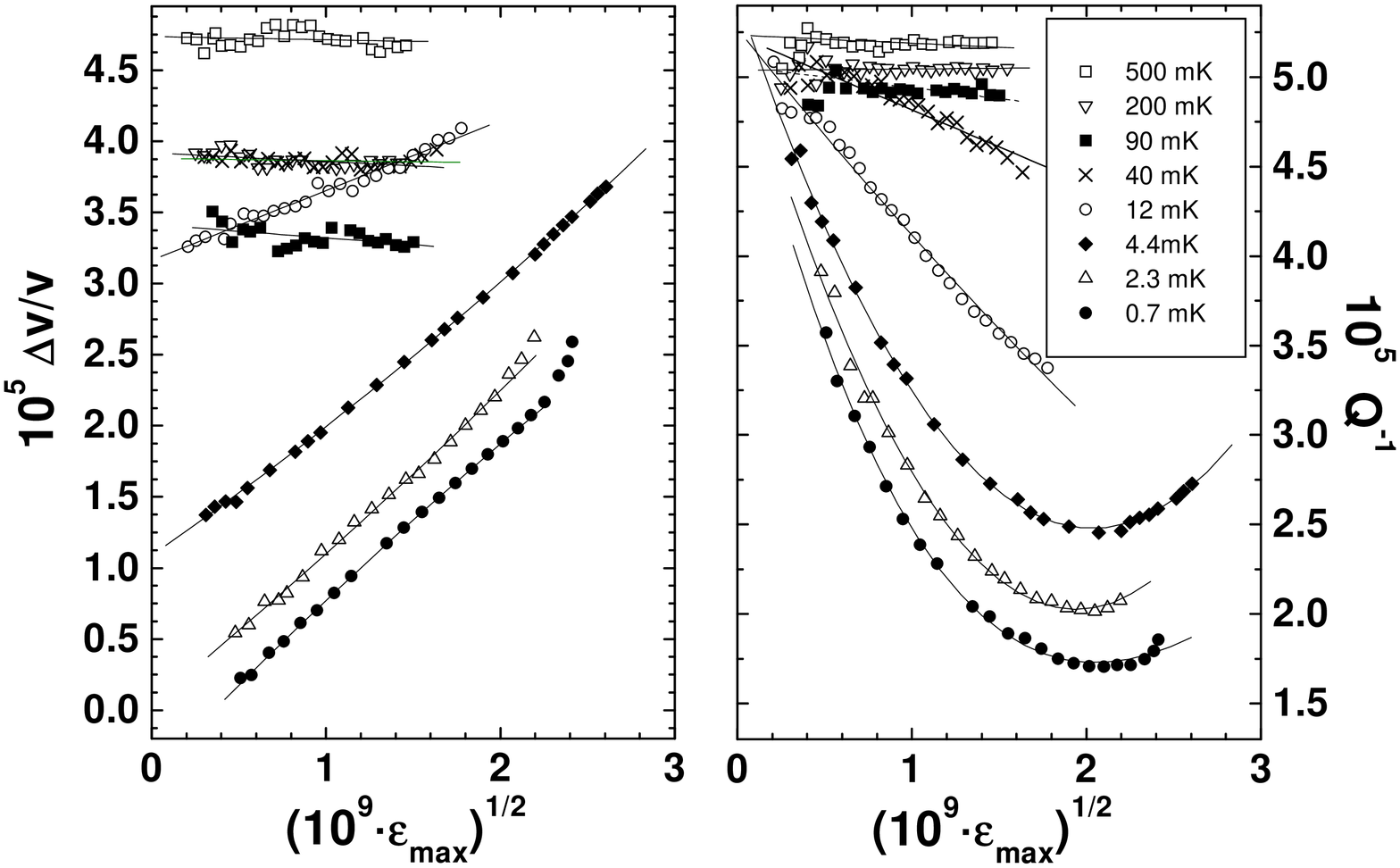}
\caption{Sound-velocity variation $\Delta v/v$ (left) and internal friction $Q^{-1}$ (right) of Zr$_{30}$Cu$_{70}$ at 1 kHz, plotted versus square-rooted maximum strain of the vibrating reed, for several series of measurements performed at selected constant temperatures (see legend). Solid lines show polynomial fits used to find zero-strain extrapolated values.}
\label{strain}
\end{figure}

However, below $\sim 10$~mK both $\Delta v/v$ and $Q^{-1}$ strongly depend on the excitation voltage (hence on strain), a behavior similar to that observed in metallic glass PdSiCu \cite{esqu92,ramos00}. Since the STM is linear in strain, it cannot account for any strain dependence of acoustic properties, with exception of the observed saturation at high intensities of resonant absorption at ultrasonic frequencies \cite{hunk86,esqu98}. To study this effect in detail, we have conducted complementary experiments by measuring, at selected constant temperatures, acoustic properties as a function of the applied strain which was obtained from the reed vibration amplitude \cite{ramos00}. This is important since constant-voltage curves do not mean constant-strain curves at very low temperature: the applied strain typically increases a factor of 3 with decreasing temperature along our constant-voltage data sets. We show in Fig.~\ref{strain} these measurements taken for Zr$_{30}$Cu$_{70}$ (no magnetic field), plotted for convenience versus square-rooted strain to stretch the lower strain region and facilitate the extrapolation of data to zero limit. Although not shown here, a very similar behavior is observed in the normal state ($B = 250$~mT) and for Zr$_{40}$Cu$_{60}$. As anticipated from Fig.~\ref{temperature}, sound velocity increases and internal friction decreases with strain, below at least 12 mK.  
In addition, at relatively high strains, a minimum value of $Q^{-1}$ is observed. This also unexpected feature will not be studied here.

In the same spirit of Ref. \cite{ramos00}, low-degree polynomial fits (see Fig.~\ref{strain}) have been used to determine zero-strain extrapolated data, which are shown in Fig.~\ref{temperature}. Although the lack of understanding of these strain-dependent effects --and hence of a function with physical meaning-- hinders a proper determination of zero-strain values (as already noticed in PdSiCu \cite{ramos00}), some conclusions can be drawn: (i) sound velocity seems to recover the STM behavior in the zero-strain limit, though still exhibiting a small saturation below a few mK, and with a logarithmic slope in the SC state approximately double of that of the N state, as predicted by the STM; (ii) in the N state, $Q^{-1}$ also seems to recover the expected plateau (as observed in normal metallic PdSiCu \cite{ramos00}); (iii) nonetheless, this reappareance of the plateau in $Q^{-1}$ down to the lowest temperatures occurs in the SC state too (instead of $\sim T^3$), an absolutely unexpected result. Therefore, the known observation in SC glasses \cite{hunk86,esqu98,rayc84} of a decrease in the internal friction below the plateau value at lower temperatures is shown here to be due to the applied strain, not because of the expected "crossover" at $T_{\rm co} \approx 30$~mK. 

\begin{figure}[t]
\includegraphics[width=10cm,angle=270,clip]{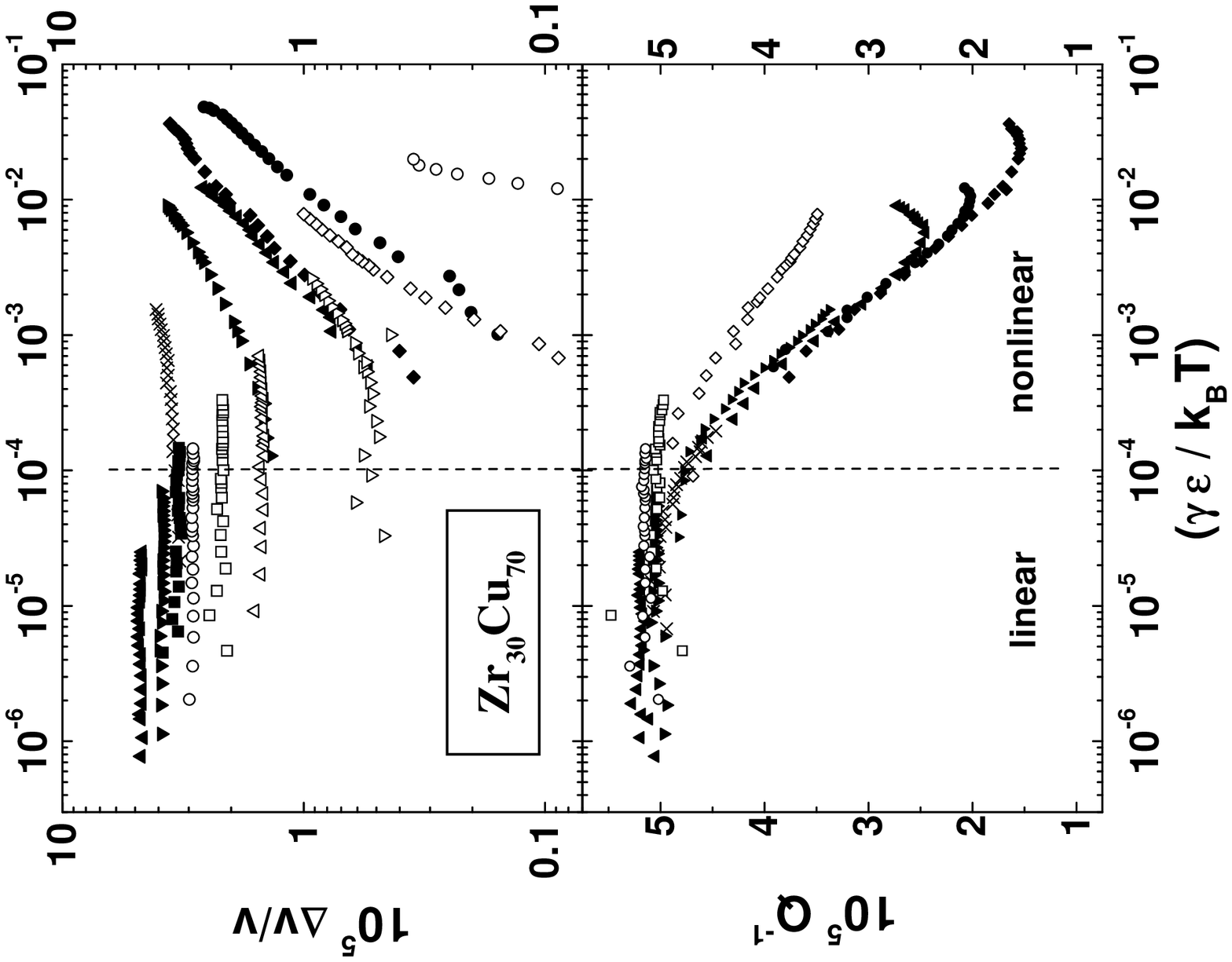}
\caption{Sound-velocity variation $\Delta v/v$ (top) and internal friction $Q^{-1}$ (bottom) measured for Zr$_{30}$Cu$_{70}$ (solid symbols, $B = 0$; open symbols, $B = 250$ mT), plotted versus the ratio of strain energy to thermal energy, for several sets of constant-temperature data such as those in Fig. \ref{strain}.}
\label{masterplot}
\end{figure}

The search for a deeper comprehension of these strain-dependent effects on acoustic properties, as well as the need of a meaningful physical function to follow their behavior at very low strain, led us to find out the striking representation shown in Fig.~\ref{masterplot}. All sets of constant-temperature $Q^{-1}$ data merge into a common curve, resembling tanh($1/x$), within experimental error when displayed versus the dimensionless ratio of strain energy to thermal energy along almost 5 orders of magnitude (with the obvious exception of the abovementioned temperature-independent upturn at higher strains, irrelevant for the present discussion). Corresponding curves measured in the normal state (so expected to be dominated by electronic instead of by phononic relaxation) are also shown there. A typical value of $\gamma = 1 $~eV has been used for the coupling constant. As can be observed, all curves show the same $Q^{-1}$ ``plateau'' value at $\gamma \epsilon / k_{\rm B} T \, < \, 10^{-4}$ (strikingly, also in the SC, dielectric case). Moderately increasing the strain, so that $\gamma \epsilon / k_{\rm B} T \, > \, 10^{-4}$, makes $Q^{-1}$ to decrease with a roughly logarithmic dependence on $\gamma \epsilon / k_{\rm B} T $. When phonons are replaced by electrons as TS relaxation centers in the normal state, the same behavior is observed, though the slope is less steep. The problem encountered in zero-strain extrapolations at the lowest temperatures (say, below 5--10 mK) is also made clearer now: even the very low strains used here are not low enough to completely reach the ``plateau''  limit. On the other hand, $\Delta v/v$, which is much more governed by resonant processes (especially in the SC state), also exhibit a crossover at $\gamma \epsilon / k_{\rm B} T \, \sim \, 10^{-4}$ from the strain-independent behavior expected within the STM into a regime of strong increase of the sound velocity with strain, either in N or SC states. 

\begin{figure}[htb]
\includegraphics[width=5cm,angle=270,clip]{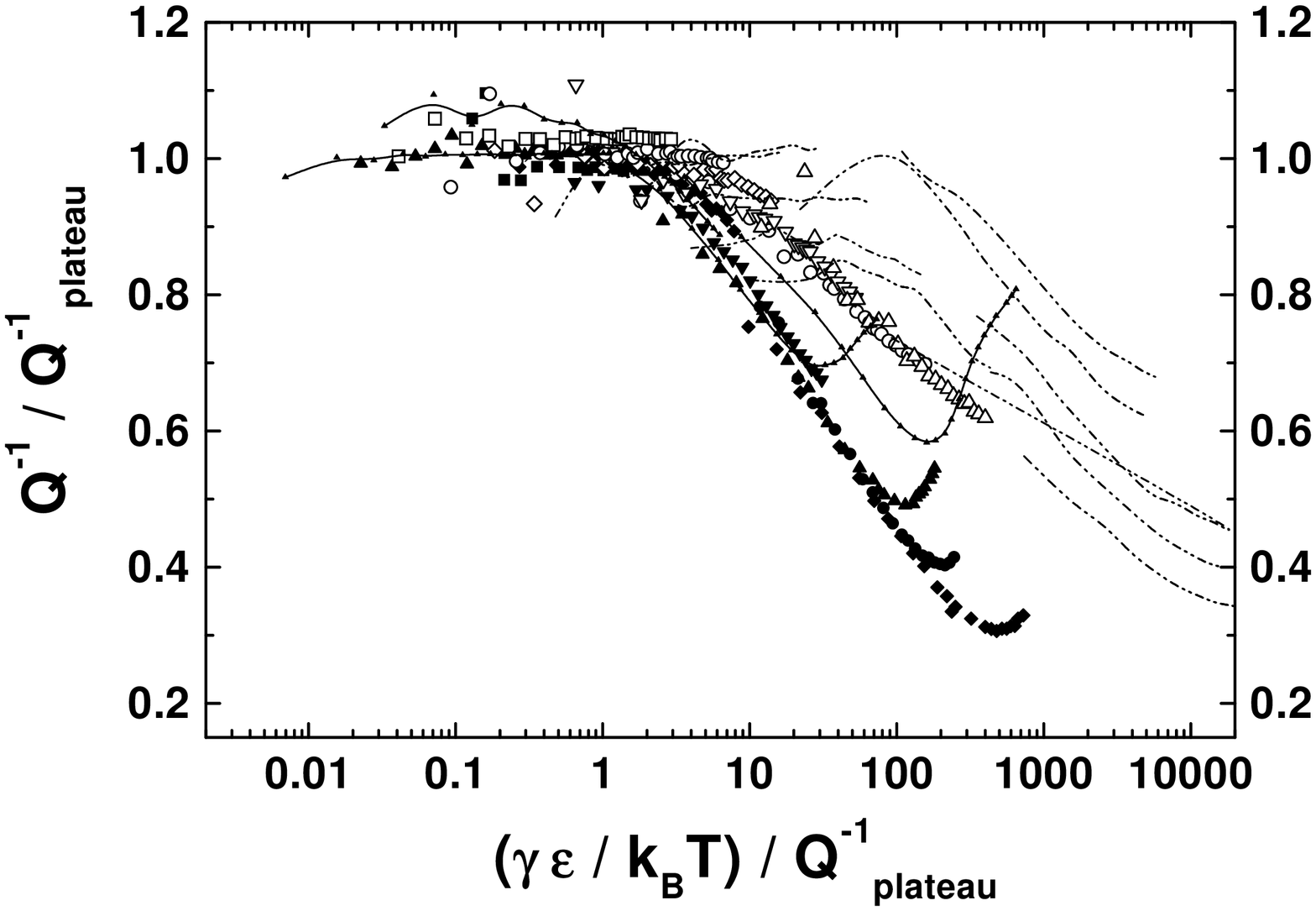}
\caption{Internal friction normalized to its plateau value $Q^{-1} / Q^{-1}_{\rm plateau}$ versus the ratio of strain to thermal energies in terms of $Q^{-1}_{\rm plateau}$ from different sets of measurements: Zr$_{30}$Cu$_{70}$ (solid symbols, $B = 0$; open symbols, $B = 250$ mT), Zr$_{40}$Cu$_{60}$ (solid lines, $B = 0$), normal conducting
PdSiCu (dashed lines, data from {\protect\cite{ramos00}}). }
\label{universal}
\end{figure}
                                                                                         
Furthermore, although we have shown and discussed up to now only the case of Zr$_{30}$Cu$_{70}$ for sake of clarity, a very similar plot can be obtained for Zr$_{40}$Cu$_{60}$ data in the SC state, and also when we use earlier published data \cite{ramos00} of the normal metal PdSiCu, pointing once again to a universal behavior of glasses.
In Fig.~\ref{universal}, internal friction data for all these cases have been normalized to their corresponding constant value found at the plateau $Q^{-1}_{\rm plateau}$, assuming always $\gamma = 1 $~eV. Indeed, the reason for this normalization of $\gamma \epsilon / k_{\rm B} T$ in terms of  $Q^{-1}_{\rm plateau}$ is to emphasize that in all cases the crossover from strain-independent behavior to strain-dependent one occurs when $\gamma \epsilon / k_{\rm B} T \, \approx $ 2--5~$Q^{-1}_{\rm plateau} \approx  P_0  U_0$ (notice \cite{justify} as well as the uncertainty in $\gamma$). It represents the main result of this work.

We want to show now that the observed general behavior in Fig.~\ref{universal} is in quantitative agreement with
the model of interacting TS \cite{yu} and that our strain-energy scans at constant temperatures enable us to assess directly (within the uncertainty in $\gamma$) the interaction strength between TS.
We note first that the observed behavior cannot be ascribed to the saturation effects proposed by Stockburger {\it et al.} \cite{stockb}, since the saturation regime should be reached when $\gamma \epsilon / k_{\rm B} T \sim 1$, i.e. at
4 orders-of-magnitude higher strains.             
In short, TS pairs can interact through the exchange of virtual phonons. $P_0 \times k_{\rm B} T$  is the number of thermally activated TS per unit volume, and then $P_0 U_0 \times k_{\rm B} T$ is the total effective interaction energy between TS at the considered temperature. If the strain energy driven by the sound wave $\gamma \epsilon$ is less than that interaction energy, one may observe the ``true'' unperturbed response of the TS, in which only their interaction and the thermal energy determine the temperature dependence of the acoustic properties. However, when $\gamma \epsilon \, > \, ( P_0 U_0 \times k_{\rm B} T )$, the system is perturbed and nonlinear effects will affect the acoustic properties,  the STM picture of independent TS being no longer valid. 
The dashed line in Fig.~\ref{masterplot} would therefore indicate the border between the {\em independent}-TS, linear regime (left) and the nonlinear regime arising from TS interaction at high enough strain/temperature ratios (right). This crossover will occur around $\sim $ 20--50~mK for typical experiments as already mentioned, but strictly speaking depends on the variable $\epsilon / T$, since strain and thermal energies play an inextricably combined role in the acoustic properties of glasses at low temperatures due to the mutual interaction of TS. The challenge remains for theoreticians to account in detail for the observed behavior in the nonlinear regime (a logarithmic decrease of sound absorption and a corresponding increase of sound velocity with the strain --in terms of TS thermal energy--), as well as for the specific behavior due to TS relaxational processes (at least when mediated by phonons) even in the linear regime, which has been shown to clearly deviate from the STM predictions too. 

In summary, acoustic experiments in metallic glasses, both in SC and N states, with variation of the strain energy used, have enabled us to obtain the interaction strength between TS in amorphous solids which is found to be in quantitative agreement with the model of interacting TS \cite{yu}. Our results provide a clue to generally account for the overall observed experimental deviations of the acoustic properties in glasses from the predictions of the STM. We have been able to study in detail this behavior by using the vibrating-reed technique which, in contrast to other techniques, allows measuring at very low strains only limited by experimental resolution. 

This work was supported by the ``TMR-Large Scale Facility Program: contract No. ERB FMGE CT950072'' of the EU.
Valuable discussions with A. Burin and many other participants of the Workshop on Collective Phenomena in the
Low-Temperature Physics of Glasses (Dresden, 2000), organized by S. Hunklinger, C. Enss and R. K\"uhn, are gratefully acknowledged.

\end{document}